\documentclass[%
 reprint,
superscriptaddress,
 amsmath,amssymb,
 aps,
]{revtex4-2}

\usepackage{graphicx}
\usepackage{dcolumn}
\usepackage{bm}
\usepackage{hyperref}
\usepackage[dvipsnames]{xcolor}
\usepackage{soul}

\begin{document}

\preprint{APS/123-QED}

\title{Uniaxial strain effects on the Fermi surface and quantum mobility of the Dirac nodal-line semimetal ZrSiS}

\author{J. P. Lorenz}
 \email{j.p.lorenz@uva.nl}
 \affiliation{Van der Waals-Zeeman Institute, University of Amsterdam, Science Park 904, 1098 XH Amsterdam, The Netherlands}
\author{J. F. Linnartz}
 \affiliation{High Field Magnet Laboratory (HMFL-EMFL) \& Institute for Molecules and Materials, Radboud University, Toernooiveld 7, 6525 ED Nijmegen, The Netherlands}
\author{A. Kool}
 \affiliation{High Field Magnet Laboratory (HMFL-EMFL) \& Institute for Molecules and Materials, Radboud University, Toernooiveld 7, 6525 ED Nijmegen, The Netherlands}
\author{M. R. van Delft}
 \affiliation{High Field Magnet Laboratory (HMFL-EMFL) \& Institute for Molecules and Materials, Radboud University, Toernooiveld 7, 6525 ED Nijmegen, The Netherlands}
\author{W. Guo}
 \affiliation{Institute for Theoretical Physics Amsterdam, University of Amsterdam and European Theoretical Spectroscopy Facility, Science Park 904, 1098 XH Amsterdam, The Netherlands}
\author{I. Aguilera}
 \affiliation{Institute for Theoretical Physics Amsterdam, University of Amsterdam and European Theoretical Spectroscopy Facility, Science Park 904, 1098 XH Amsterdam, The Netherlands}
\author{R. Singha}
 \affiliation{Department of Chemistry, Princeton University, \\Princeton, New Jersey 08544, USA}
\author{L. M. Schoop}
 \affiliation{Department of Chemistry, Princeton University, \\Princeton, New Jersey 08544, USA}
\author{N. E. Hussey}
 \affiliation{High Field Magnet Laboratory (HMFL-EMFL) \& Institute for Molecules and Materials, Radboud University, Toernooiveld 7, 6525 ED Nijmegen, The Netherlands}
 \affiliation{H. H. Wills Physics Laboratory, University of Bristol, \\Tyndall Avenue, BS8 1TL, United Kingdom}
\author{S. Wiedmann}
 \affiliation{High Field Magnet Laboratory (HMFL-EMFL) \& Institute for Molecules and Materials, Radboud University, Toernooiveld 7, 6525 ED Nijmegen, The Netherlands}
\author{A. de Visser}
 \email{a.devisser@uva.nl}
 \affiliation{Van der Waals-Zeeman Institute, University of Amsterdam, Science Park 904, 1098 XH Amsterdam, The Netherlands}%

\date{\today}       

\begin{abstract}
ZrSiS has been identified as an exemplary Dirac nodal-line semimetal, in which the Dirac band crossings extend along a closed loop in momentum space. Recently, the topology of the Fermi surface of ZrSiS was uncovered in great detail by quantum oscillation studies. For a magnetic field along the tetragonal $c$~axis, a rich frequency spectrum was observed stemming from the principal electron and hole pockets, and multiple magnetic breakdown orbits.  In this work we use uniaxial strain as a tuning parameter for the Fermi surface and the low energy excitations. We measure the magnetoresistance of a single crystal under tensile (up to $0.34~\%$) and compressive (up to $-0.28~\%$) strain exerted along the $a$~axis and in magnetic fields up to 30~T. We observe a systematic weakening of the peak structure in the Shubnikov-de Haas frequency spectrum upon changing from compressive to tensile strain. This effect may be explained by a decrease in the effective quantum mobility upon decreasing the $c/a$ ratio, which is corroborated by a concurrent increase in the Dingle temperature.
\end{abstract}

\maketitle

\section{\label{sec:level1}Introduction}
When investigating the correlated and topological properties of crystalline materials their Fermi surface is often studied, because knowledge of the low-energy excitations close to the Fermi level can foster a fundamental understanding of the electronic physics at play. In the case of topological matter, Dirac nodal-line semimetals (NLSMs) provide an excellent playground \cite{Weng15, Wu16, Bian16, Neup16, Scho16}. In these systems the valence and conduction bands cross each other in a closed loop (the nodal line) inside the Brillouin zone. The topological character of these materials has been established in quantum oscillation experiments \cite{Wang16, Ali16, Hu2016, Sing17, Hu17}. The small density of states near the Fermi level can help shine light on the correlated character, as this is predicted to reduce the screening of the long-ranged Coulomb interaction \cite{Huh16}, making NLSMs more susceptible to various types of order such as superconductivity or magnetism \cite{Liu17, Roy17, Rud18}.

NLSMs have been experimentally investigated in the recent past, for instance PtSn$_4$ and PbTaSe$_2$ \cite{Wu16,Bian16}. In these two materials the Fermi level is shared by a nodal line and topologically trivial bands. The latter makes it more challenging to search for correlation effects in a topological material, as the quasiparticle behaviour of the crystal is not governed solely by the electronic states on the nodal line. Conversely, in the NLSM ZrSiS the nodal line is the only band feature near its Fermi level. The dispersion of the bands extends linearly for a relatively large energy range (0--2~eV), with only a small gap ($\sim0.02$~eV) in the Dirac spectrum as a result of spin-orbit coupling. This gives rise to a cage-like 3D Fermi surface and results in the physical behaviour of ZrSiS being governed by practically only the topological aspects of its electronic structure \cite{Neup16}. This makes ZrSiS a very appealing choice for studying correlated topological matter.

In previous research ZrSiS was studied by means of quantum oscillations (QOs), identifying its extremal Fermi surface cross sections \cite{Wang16, Ali16,Sing17,Hu17}. In the $Z$-$R$-$A$ plane, these are the fundamental $\alpha$ and $\beta$ pockets. Subsequent work also includes QOs as a result of magnetic breakdown (MB), and compares them to density functional theory (DFT) calculations \cite{Pezz18,Delft19,Mull20}. MB occurs in ZrSiS when a sufficiently high magnetic field causes quasiparticles to tunnel between the different fundamental pockets on the nodal line, overcoming the small gap introduced by spin-orbit coupling \cite{Shoenberg}. In ZrSiS, this gap (and the nodal-line state) is topologically protected by both the mirror- and inversion symmetry of the crystal structure \cite{Gu19}. Recently, the physical behaviour of ZrSiS has been probed using hydrostatic pressure as a tuning parameter with pressures as high as 57~GPa \cite{Sing18}. The observed changes in the Fermi surface properties suggests the presence of a pressure-driven topological quantum phase transition \cite{Sing18,Gu19,VanG19}. Closely related to hydrostatic pressure, another powerful experimental tool for exploration of the topological properties of band structures is the application of uniaxial pressure. In the elastic regime of the sample, the applied stress relates linearly to the resulting strain. This uniaxial strain response of materials that have exceptional topological band-structure features has been of great interest in the past few years. Strain has been used to demonstrate tunability of Dirac states \cite{Liu14, Heui14}, shift phase transition temperatures of materials \cite{Step17, Ike18, Mali20}, induce topological phase transitions \cite{Wang17, Step17, Mut19}, and lift the degeneracy of Fermi surface pockets \cite{Schi20}. As regards ZrSiS, electronic structure calculations have predicted significant effects of uniaxial strain on the Fermi surface, notably a large increase of the MB gap.~\cite{Zhou20} These calculations have been made for tensile strains between 1 and $4~\%$, but only for strains along the $c$~axis. We remark, such high strain values are difficult to realize experimentally.

In this paper, we report on the effect of uniaxial strain on the Fermi surface of ZrSiS. A single crystal of ZrSiS was mounted onto a Razorbill CS100 cryostrain cell and its magnetoresistance (MR) was measured as a function of temperature under uniaxial stress applied along the $a$~axis in magnetic fields up to 30~T. Together with a related study on ZrSiSe \cite{Linnartz2023}, this is the first experiment in which a CS100 strain cell was used to measure a sample at fields this high. Strain-induced changes in the Shubnikov-de Haas (SdH) oscillations present in the MR data were analyzed using fast Fourier transforms (FFTs) and the resulting frequency spectra at each temperature- and strain value were compared. We clearly identify the $\beta$~peak at 418~T, which has been absent in previous SdH studies \cite{Pezz18,Mull20}. Moreover, the contribution of the fundamental $\beta$~orbit to the conductivity of the sample relative to the contribution of the $\alpha$~orbit is found to be enhanced in the uniaxially compressed state. Compressive strain also increases the FFT amplitudes of the MB peaks, with some of the linear combinations of fundamental frequencies only being found in the data of the compressed state. In each case, tensile strain demonstrates the opposite effect. DFT-based band structure calculations for strain along the $a$~axis corroborate our results.

\section{Experiment}
ZrSiS single crystals were grown in a carbon-coated quartz tube using stochiometric amounts of each element together with a small amount of I$_2$. The tube was then vacuum sealed and heated to 1100~$^{\circ}$C for one week. The applied temperature gradient was 100~$^{\circ}$C. After extracting the crystals they were wrapped in Zr~foil and annealed under high vacuum at 600~$^{\circ}$C for three weeks. The crystal structure and composition were verified using powder X-ray diffraction and energy-dispersive X-ray spectroscopy (EDX).

The crystals were cut into small bar-shaped samples with dimensions of 1--1.5~$\times$ 0.2~$\times$~0.1~mm$^3$ ($a$~$\times$~$b$~$\times$~$c$~axis). Thin titanium and gold layers were evaporated on four lines across the samples in order to create proper contact pads for the wiring. A sample was mounted on the strain cell (CS100, Razorbill Instruments Ltd \cite{cs1x0, hicks14}) using Araldite glue, in a way that uniaxial stress could be applied along the $a$~axis and a magnetic field could be applied along the $c$~axis. For a detailed description of the mounted cell set-up see Fig.~S1 of the Supplemental Material (SM) \cite{supp}. Four 25~$\mu$m gold wires were arranged according to the four-point probe method to measure the resistance with a current along the $a$~axis. The wires were connected to the contact pads on the sample using DuPont 4929N silver conductive paint. Sample resistance was measured with a Stanford Research Systems SR865A Lock-In amplifier at a constant current excitation of 2~mA. The uniaxially applied stress was measured via a precalibrated capacitive sensor inside the strain cell using an Andeen-Hagerling 2700A capacitance bridge. The uniaxial strain $\varepsilon_{\textrm{disp}}~=~\Delta L / L$ experienced by the sample was then derived from the applied displacement $\Delta L$ (a few $\mu$m) and the shortest distance $L$ between the two glued ends of the sample (see Fig.~S1 of the SM \cite{supp}). In-depth information on strain homogeneity, limitations and reading errors can be found in the Razorbill CS1x0 Product Datasheet \cite{cs1x0}.

\nocite{Jong15,Friedrich2010,Perdew1996,wannier90,MatProj2013}

All transport measurements were performed in a $^4$He bath cryostat with a base temperature of 1.4~K. The magnetic field sweeps were carried out using a resistive Bitter magnet at the High Field Magnet Laboratory (HFML) at Radboud University with the magnetic field direction along the $c$~axis of the crystal.

\section{Results and Analysis}
\begin{figure}[b]
\includegraphics[scale=0.21]{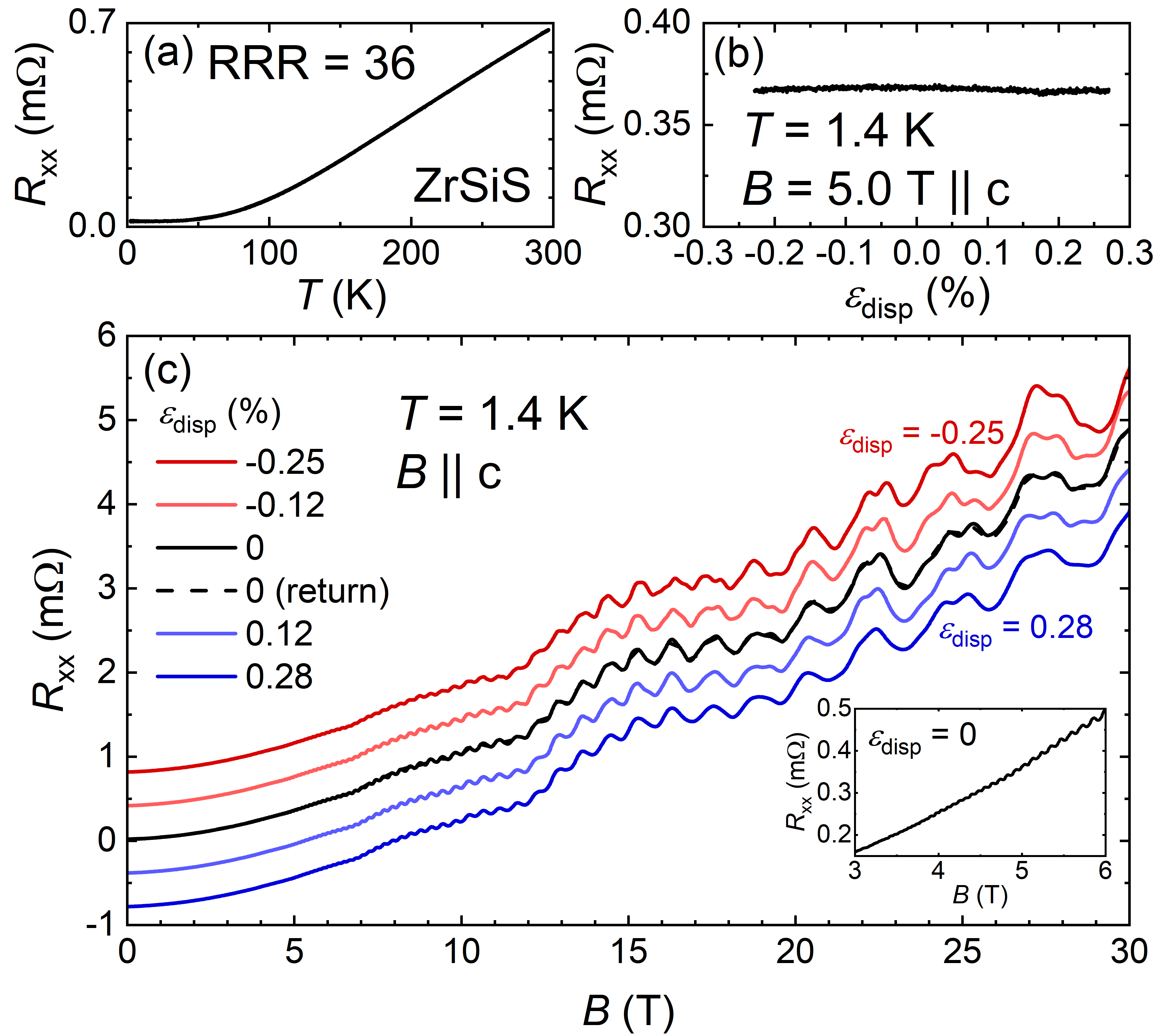} 
\caption{(a) Cooldown $R(T)$ curve. The longitudinal resistance $R_{\textrm{xx}}$ in m$\Omega$ of the ZrSiS sample against temperature $T$ in K. The residual resistance ratio of the sample is 36. (b) Longitudinal resistance $R_{\textrm{xx}}$ in m$\Omega$ against the applied uniaxial strain $\varepsilon_{{\textrm{disp}}}$ along the $a$~axis~[100], at 1.4~K and 5~T. Negative strain means compression, positive strain means tension. (c) Longitudinal resistance $R_{\textrm{xx}}$ in m$\Omega$ against the applied magnetic field $B$ in Tesla for negative and positive strain at 1.4~K. The curves have been offset for clarity, with the strain for each curve going from negative to positive in a descending fashion. A final zero strain measurement (dashed black line) falls onto the solid black line and verifies having stayed within the elastic regime. The inset shows SdH oscillations of the zero strain measurement being present at $B$~=~4~T.}
\label{fig:rawdata}
\end{figure}

\begin{figure*}
\includegraphics[scale=0.64]{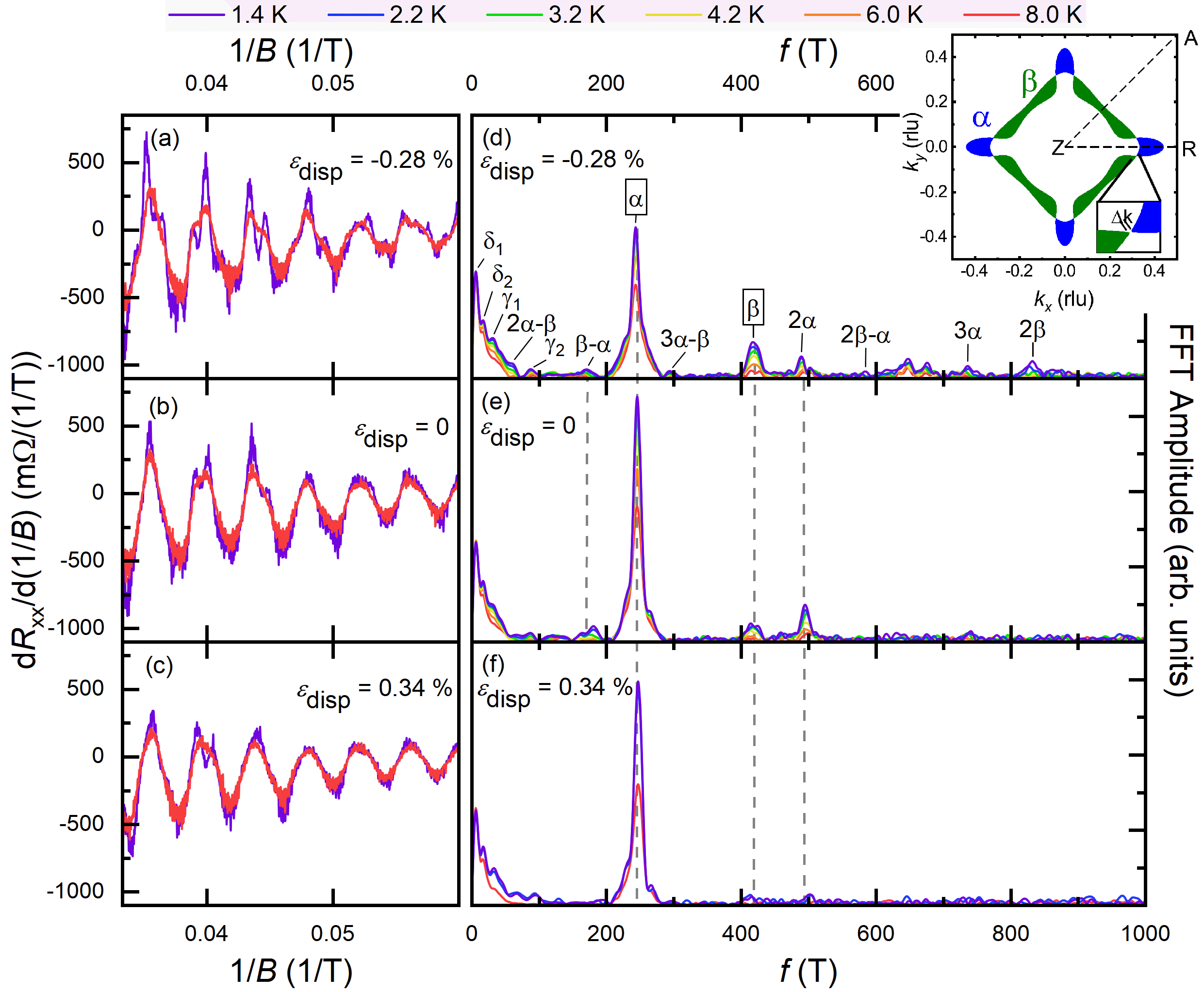}
\caption{(a--c) Derivative plots of $R_{\textrm{xx}}$ against 1/$B$ at varying $T$, for $0.28~\%$ compression, zero strain and $0.34~\%$ tension. Only the $T$ curves 1.4~K and 8~K are shown here for clarity. (d--f) FFT plots of the derivative data at corresponding $T$ (indicated by the different colors) and strain values. The identification of the peak frequencies is given in (d). Vertical dashed lines link peaks of orbits in the $Z$-$R$-$A$ plane under compression to those at different strain values, if visible. The inset shows the two distinct electron (green, $\beta$) and hole (blue, $\alpha$) pockets in the $Z$-$R$-$A$ plane of ZrSiS, in the zero strain state. These pockets are separated by a small gap in momentum space $\Delta k$ that arises due to spin-orbit coupling. Inset figure taken from Ref.~\cite{Mull20}, courtesy of the authors.}
\label{fig:ffts}
\end{figure*}

The sample's longitudinal resistance $R_{\textrm{xx}}$ is shown against temperature $T$ in Fig.~\ref{fig:rawdata}{(a)}. This curve taken upon cooling reveals a residual resistance ratio (RRR) value of 36, which proves to be sufficiently high for observing pronounced SdH oscillations. Fig.~\ref{fig:rawdata}{(b)} shows $R_{\textrm{xx}}$ against sample strain $\varepsilon_{\textrm{disp}}$ at $T~=~1.4$~K and $B~=~5$~T. Negative strain values correspond to compression, while positive strain values correspond to tension. Uniaxial stress was applied along the $a$~axis. The data show that there is no significant change in $R_{\textrm{xx}}$ with strain. This implies that there is no elasto-magnetoresistance (EMR) present in this material under these conditions. A set of MR data is shown in Fig.~\ref{fig:rawdata}{(c)}, consisting of six individual field sweeps at different strain values along the $a$~axis (curves are offset for clarity), with $T~=~1.4$~K and $B \parallel c$. The chronological order of the sweeps starts with a zero strain measurement and is followed by the increasing compression measurements, shown in red. Then, the increasing tension measurements in blue were performed, ending with another zero strain measurement (dashed black line). Both compressive and tensile strain demonstrate a systematic evolution in curve shape with strain. The curves of the two zero strain measurements overlap well, which indicates having stayed within the elastic strain regime of the sample. In Fig.~S2 of the SM the difference in $R_{\textrm{xx}}(B)$ between the strained curves and the zero strain curve shows the tunability of the quantum oscillations through the use of uniaxial strain along the $a$~axis \cite{supp}. 

In order to investigate the temperature dependence of the MR at different strain values a second run of strain measurements was performed, the raw MR data of which are shown and discussed in Fig.~S3 of the SM \cite{supp}. In Fig.~\ref{fig:ffts}{(a--c)} the derivative of these MR with respect to the inverse magnetic field $1/B$ is shown against $1/B$ in T$^{-1}$, at the minimum ($1.4$~K) and maximum ($8.0$~K) temperature and at strain values of $\varepsilon_{\textrm{disp}}~=~-0.28~\%$, zero and $0.34~\%$. Here the other temperature curves ($2.2$~K $\leq T \leq 6.0$~K) have been omitted to more clearly demonstrate the temperature and strain effects. The FFT analysis of the corresponding derivatives at all measured $T$ is shown in Fig.~\ref{fig:ffts}{(d--f)}. The FFT magnetic field range was chosen to be $5 - 30$~T and the Hann window function was used. The frequency peaks in Fig.~\ref{fig:ffts}{(d)} have been labeled using the fundamental quasiparticle orbits $\alpha$ and $\beta$ (shown in the inset), their harmonics and their linear combinations \cite{Mull20}. Low-frequency orbits ($\delta$, $\gamma$) have also been identified, but the focus remains on the higher frequencies. Interestingly, the FFTs show a clear enhancement in frequency peak structure towards compressive strain $\varepsilon_{\textrm{disp}}~=~-0.28~\%$, whereas a tensile strain of $\varepsilon_{\textrm{disp}}~=~0.34~\%$ shows the opposite effect. Frequency peaks that are visible at multiple strain values are linked with vertical gray dashed lines. Similar features are observed in the FFTs of the derivatives of the MR data of Fig.~\ref{fig:rawdata}{c}. See Fig.~S4 of the SM \cite{supp}.

Fig.~\ref{fig:ffts} evidently shows a large change in proportion of the FFT amplitude between the two fundamental frequencies in the $Z$-$R$-$A$ plane. We also note that under a uniaxial compression of $\varepsilon_{\textrm{disp}}~=~-0.28~\%$ the $\alpha$/$\beta$ FFT amplitude ratio has a value less than half (4.5) of that compared to the zero strain case (9.8).

Next we derive the quasiparticle cyclotron mass $m_c$ of the fundamental frequencies $\alpha$ and $\beta$ for different magnetic field ranges. We use the standard thermal damping factor of the Lifshitz-Kosevich formula $R_T~=~X/\sinh{(X)}$, where $X~=~14.69 m_cT/\langle B \rangle$ \cite{Shoenberg,Kuma19}, with an average field $\langle B \rangle~=~2/(B_{min}^{-1} + B_{max}^{-1})$. The results are reported in Table~\ref{tab:emlist}. For the full range, $B=$~5~--~30~T for $\alpha$ and 10~--~30~T for $\beta$, the zero strain $m_{c,{\textrm{full range}}}$ values correspond well to those of previous research \cite{Sing17,Pezz18,Mull20}. The compressive strain mass values appear to be slightly lower than at zero strain, but given the error margin this is not a significant difference.

\begin{table}[b]
\caption{\label{tab:emlist}%
Fundamental orbits and their frequencies at zero, $-0.28~\%$ compressive and $0.34~\%$ tensile strain. Cyclotron masses in $m_e$ acquired for different magnetic field ranges. Full range is 5~--~30~T for the $\alpha$ and 10~--~30~T for the $\beta$ pocket. The $m_{c,{\textrm{varying } \langle B \rangle}}$ values are taken from the field-dependent study (see text and Fig.~\ref{fig:ems}). The entries in the rightmost column are obtained in the range 4~--~11~T and are used for the determination of $T_D$ (see text).
}
\renewcommand{\arraystretch}{1.2}
\begin{ruledtabular}
\begin{tabular}{lcccc}
\textrm{Orbit} & 
\textrm{$f$ (T)} & 
\textrm{$m_{c,{\textrm{full range}}}$} & 
\textrm{$m_{c,{\textrm{varying } \langle B \rangle}}$} &
\textrm{$m_{c,{4 - 11 \textrm{ T}}}$}\\[1mm]
\colrule\\[-3mm]
$\alpha_0$ & 245 & 0.15 $\pm$ 0.01 & 0.20 $\pm$ 0.04 & 0.16 $\pm$ 0.01\\
$\alpha_{\textrm{comp}}$ & 243 & 0.13 $\pm$ 0.01 & 0.18 $\pm$ 0.04 & 0.16 $\pm$ 0.01\\
$\alpha_{\textrm{tens}}$ & 247 & 0.15 $\pm$ 0.01 & 0.21 $\pm$ 0.02 & 0.16 $\pm$ 0.01\\
$\beta_0$ & 418 & 0.42 $\pm$ 0.05 & 0.51 $\pm$ 0.10 & --\\
$\beta_{\textrm{comp}}$ & 419 & 0.39 $\pm$ 0.02 & 0.40 $\pm$ 0.06 & --\\
$\beta_{\textrm{tens}}$ & 415 & -- & 0.53 $\pm$ 0.03 & --\\[1mm]
\end{tabular}
\end{ruledtabular}
\end{table}

\begin{figure}[b]
\includegraphics[scale=0.4]{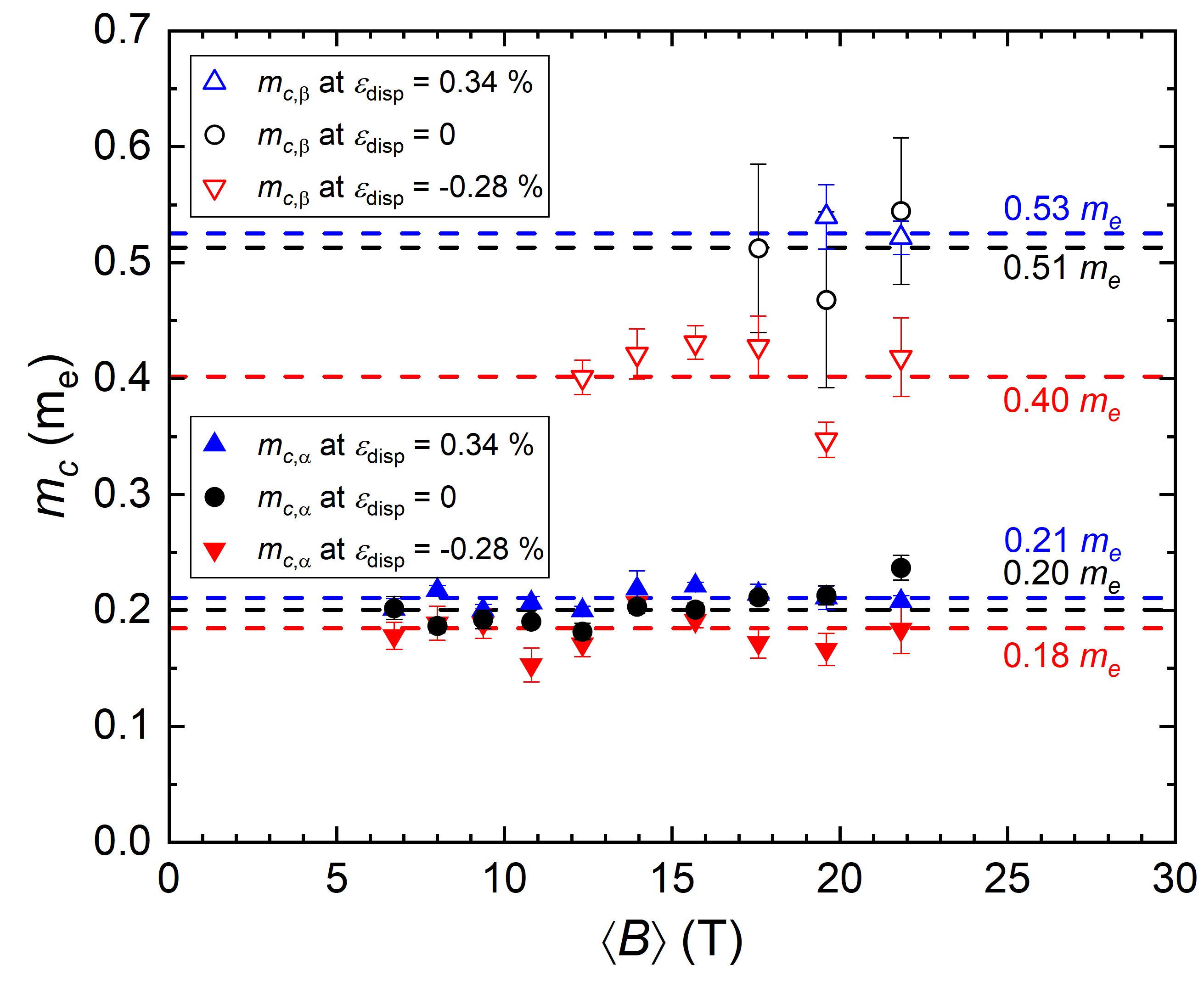}
\caption{Quasiparticle cyclotron masses of the $\alpha$ and $\beta$~orbits obtained from the $T$-dependence of the FFT amplitudes with varying average field $\langle B \rangle$. The dashed lines present the average $m_c$ values. The error bars are given by the standard error of the $m_c$ fit parameter.}
\label{fig:ems}
\end{figure}

In order to investigate the field induced mass enhancement along the nodal line reported in Ref.\cite{Pezz18}, we have calculated FFTs as described above, but at a shifting magnetic field range with a constant width $0.03174$~T$^{-1}$ in $1/B$, see subsection E of the SM \cite{supp}. The deduced cyclotron mass values for the $\alpha$ and $\beta$ orbits are plotted as a function of $\langle B \rangle$ in Fig.~\ref{fig:ems}. The data do not reveal a sizeable field dependence of $m_c$, indicating field induced correlations effects are very weak or absent. The field averaged $m_c$ values at zero, compressive and tensile strain, indicated by the horizontal dashed lines in Fig.~\ref{fig:ems}, are listed in Table~\ref{tab:emlist} as well. Again the compressive strain mass values appear to be systematically sligthly lower than at zero strain, notably for the $\beta$ orbit, but the error bar is large. We remark that the nodal-line mass enhancement reported in Ref.\cite{Pezz18} was obtained for the 600~T orbit, a frequency then attributed to the $\beta$ pocket. However, according to Ref.~\cite{Mull20} this frequency has to be attributed to the magnetic breakdown orbit $2\beta-\alpha$. The $\beta$ orbit has a frequency of 418~T (see Fig.~\ref{fig:ffts}{(d--f)}). The FFT amplitudes of the MB orbits obtained in this work are too small to perform a proper field-dependent cyclotron mass analysis.

\begin{figure}
\includegraphics[scale=0.4]{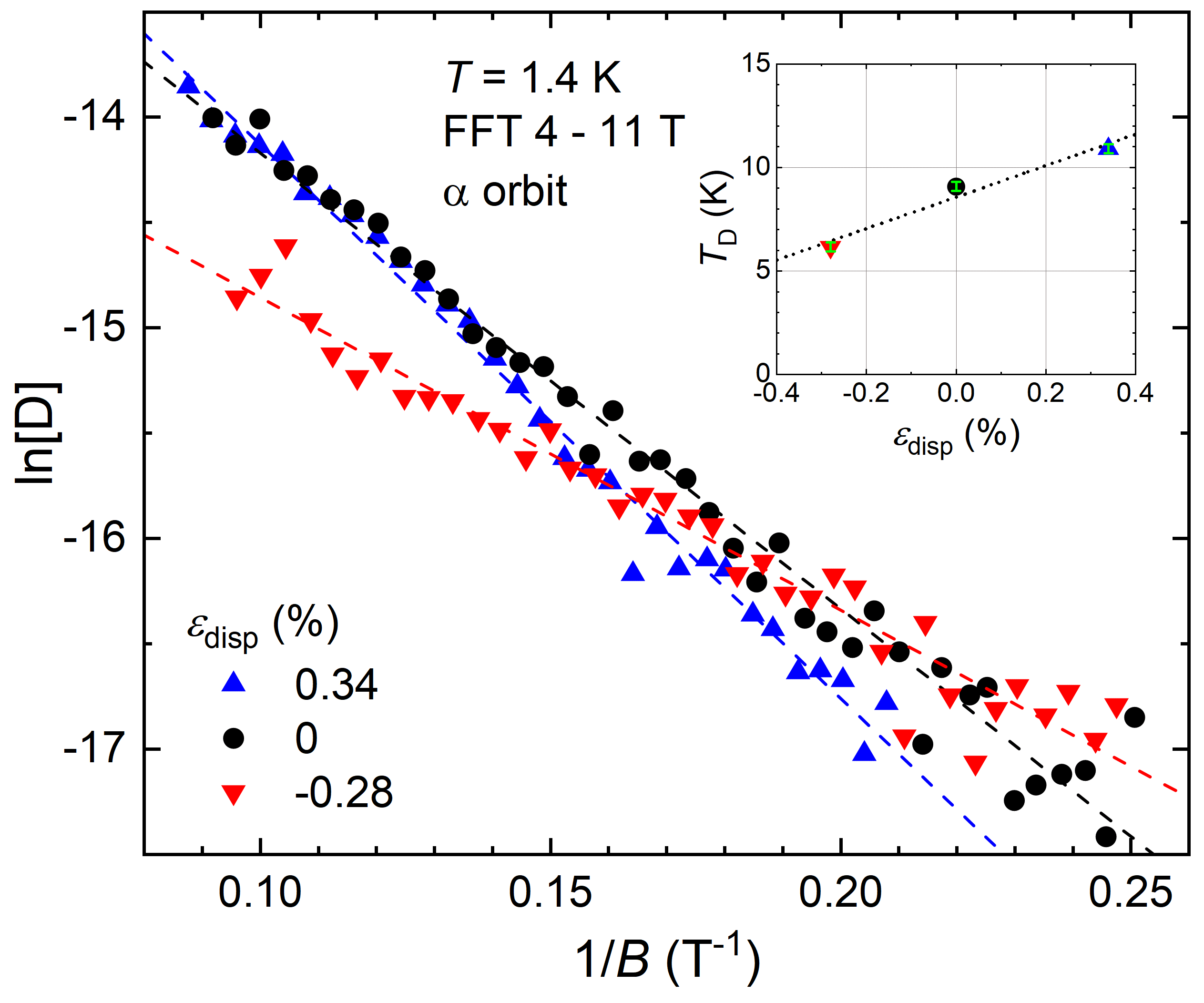}
\caption{Dingle plot of the $\alpha$~orbit at three different strain values (evaluated as in Ref.~\cite{Cao12}). ln[D] (defined in the text) is plotted as a function of 1/$B$. The Dingle temperatures $T_{\textrm{D}}$ are calculated from the slope of the linear fits (dashed lines). Inset: $T_{\textrm{D}}$ as a function of $\varepsilon_{disp} $. The error bars are small and shown in green. The dotted line is a linear fit.}
\label{fig:dingle}
\end{figure}

In order to obtain the Dingle temperature $T_{\textrm{D}}$ for the $\alpha$~orbit we show in Fig.~\ref{fig:dingle} the semilog plot of $\textrm{D}~=~\Delta RB\sinh(\xi T/\Delta E_N)$ against $1/B$ at $T~=~1.4$~K and different strain values (see Ref.~\cite{Cao12}). Here, $\Delta R$ is the SdH oscillation amplitude of the Landau level (LL) at the corresponding $B$-value, calculated from $R_{\textrm{xx}}$ by subtracting a background determined by a simple second-order polynomial fit. LLs between 4 and 11~T were used, as in this field range the $\alpha$~orbit almost completely dominates the oscillation amplitude. $\Delta E_N(B)~=~heB/2\pi m_c$ is the energy gap between the $N$th and $(N + 1)$th LL, where $m_c$ is the cyclotron mass for the $\alpha$~orbit at a field range of $4$ to $11$~T (listed in the fifth column of Table~\ref{tab:emlist}), $e$ is the electron charge and $h$ is the Planck constant. $\xi$ is a constant defined as $\xi~=~2\pi ^2 k_B$ with $k_B$ being Boltzmann's constant. The $T_{\textrm{D}}$ for each strain value can be directly calculated from the slope of the corresponding linear fit to the data \cite{Shoenberg,Cao12}. The inset of Fig.~\ref{fig:dingle} reports the obtained $T_{\textrm{D}}$ values of $6.2$~K under compression, $9.1$~K at zero strain and $10.9$~K under tension. $T_{\textrm{D}}$ can be used to acquire a quantitative estimate of the quantum mobility of the $\alpha$~orbit charge carriers, using the relation $\mu_q~=~(e\hbar /2\pi k_Bm_cT_{\textrm{D}})$ \cite{Sing17}. This gives a $\mu_q$ value of $2.1 \times 10^3$, $1.5 \times 10^3$ and $1.2 \times 10^3$~cm$^2$V$^{-1}$s$^{-1}$ for compressive, zero and tensile strain, respectively.

\section{Discussion}
This work reports the first results of uniaxially strained ZrSiS at fields up to 30~T. The identified frequency peaks in Fig.~\ref{fig:ffts}{(d--f)} can be compared to the FFT spectrum in the literature. The observed fundamental frequencies $\alpha$ and $\beta$ ($f~=~245$ and $418$~T respectively) correspond well to the literature values (see Table S1 of the SM \cite{supp}). There is no significant frequency shift as a result of uniaxial strain (see Table~\ref{tab:emlist}). Using the Onsager relation \cite{Shoenberg} $F~=~(\Phi_0/2\pi^2)A_{\textrm{ext}}$, this implies no significant change in area of any of the measured extremal Fermi surface cross sections. Magnetic breakdown (MB) peaks are situated at similar frequencies as in de Haas-van Alphen (dHvA) oscillation data analyzed in previous work \cite{Mull20}. 

A remarkable feature of our SdH data is the presence of the $\beta$~peak at $418$~T, which was not observed in previous transport studies \cite{Pezz18,Mull20}. It is clearly visible in the zero-strain FFT spectrum, and its amplitude increases under compression. This indicates an increase in the quantum mobility. Note that the $\beta$ orbit has been observed in a dHvA study by measuring torque \cite{Mull20}.

MB peaks also are more significant under uniaxial compression: $2\alpha - \beta$ and $3\alpha - \beta$ are not visible in the zero-strain and uniaxially tensioned state. In fact, tensile strain has a completely opposite effect. It suppresses all frequency peaks consisting of orbits that lie on the Dirac nodal line, implying a decrease in quantum mobility for these orbits. This decrease is corroborated by the Dingle temperatures for the $\alpha$~orbit, with $T_{\textrm{D}}$ increasing with tensile strain and decreasing with compressive strain.

$T_D$ and the quantum mobility are usually determined by the impurity and crystal defect concentration. We do not expect the change of $T_D$ is due to a change in crystal defects, since our measurements are performed in the elastic regime. However, in a layered material under strain or hydrostatic pressure the quantum mobility will depend on the $c/a$~ratio of the crystal lattice. With the distance between the layers decreasing, the inter-layer scattering will increase, which will result in a lower in-plane carrier mobility. For ZrSiS this was recently demonstrated by combined magnetotransport and X-ray diffraction measurements under hydrostatic pressure~\cite{Gu19}. In this study, the quantum mobility of the $\alpha$~orbit (designated $\beta$ in ~\cite{Gu19}) was reported to show an initial decrease at a rate of $30 $~cm$^2$V$^{-1}$s$^{-1}$/GPa. At the same time the $c/a$~ratio decreases by 0.0005~/GPa. In the case of uniaxial pressure applied along the $a$~axis we may consider $c/\sqrt{ab}$, rather than $c/a$. Using Poisson's ratio to calculate the lattice parameters under strain (see subsection G of the SM \cite{supp}) we find that $c/\sqrt{ab}$ amounts to 2.277, 2.272 and 2.267 at $-0.28~\%$ compressive, zero and $0.34~\%$ tensile strain, respectively. The corresponding values of $\mu_q$ are $2.1 \times 10^3$, $1.5 \times 10^3$ and $1.2 \times 10^3$~cm$^2$V$^{-1}$s$^{-1}$, calculated using the Dingle temperature as shown in the last paragraph of the Results section. The overall decrease of $c/\sqrt{ab}$ and $\mu_q$ matches the observed changes in the hydrostatic pressure experiment. Thus we ascertain the increase in quantum mobility can directly be linked to the increase of the $c/a$~ratio.

Another interpretation of the increase in prominence of the MB peaks under uniaxial compression is a decrease in the MB gap size caused by strain. As the MB gap size decreases, an exponentially smaller $B$ is needed in order to cause tunneling events \cite{Shoenberg}. This will lead to a lower onset magnetic field for the MB orbits and should, therefore, be visible in the FFT spectra. However, within our experimental resolution we did not detect any differences in the onset field. This is supported by DFT-based band structure calculations that we carried out at zero, compressive and tensile strain applied along the $a$ axis, see subsection H of the SM. The main results are presented in Fig.~S7, where the Fermi surface and the magnetic breakdown gap are presented at different strain values. Notably in Fig.~S7(b), it is shown that at these low strain levels significant modifications of the Fermi surface and size of the MB gap are absent.

Finally, we remark that quasiparticle orbits that go around the entire nodal loop are not observed. These orbits are referred to as full MB orbits and, in the $Z$-$R$-$A$~plane of ZrSiS, consist of eight tunneling events~\cite{Pezz18}. Full MB has been clearly observed in previous work \cite{Pezz18,Mull20} and gives rise to high frequency quantum oscillations ($f \sim 8000$~T).

\section{Summary}
To summarize, we have presented a comprehensive uniaxial strain study of ZrSiS at low temperatures and at fields up to 30~T. We measured the resistance and analyzed the SdH oscillations to find frequency peaks of extremal Fermi surface cross sections in the FFT spectra that correspond well to literature values \cite{Mull20}. The fundamental $\beta$~peak in our work is more prominent than in previous SdH research \cite{Pezz18, Mull20, Delft19}, and is found to be enhanced relative to the $\alpha$~peak in the uniaxially compressed state. In this state, the MB peaks also are more significant, showing certain linear combinations of quasiparticle orbits not visible in the zero-strain and uniaxially tensioned state. Contrary to compressive strain, tensile strain suppresses all frequency peaks consisting of orbits that lie in the $Z$-$R$-$A$~plane, implying a decrease in quantum mobility for these orbits. This decrease is supported by the calculated Dingle temperatures for the $\alpha$~orbit and attributed to a lowering of the $c/a$ ratio. DFT-based band structure calculations under strain along the $a$ axis substantiate that the FFT frequencies of the $\alpha$, $\beta$ and MB orbits do not change at these low strain values. Therefore, future uniaxial strain research on the Dirac nodal-line semimetal ZrSiS should be directed towards the application of considerable larger strain values: $|\varepsilon_{\textrm{disp}}| \gg 0.3~\%$. We note that a related strain study has recently been performed on ZrSiSe (Ref.~\cite{Linnartz2023}).

\begin{acknowledgments}
We acknowledge the support of the HFML-RU/NWO-I, member of the European Magnetic Field Laboratory (EMFL). This work was supported by the Dutch Research Council (NWO) as part of the TopCore consortium (OCENW.GROOT.2019.048). This work used the Dutch national e-infrastructure with the support of the SURF Cooperative using grants no. EINF-5380 and EINF-7851. L.M.S. was supported by the Princeton Center for Complex Materials, a National Science Foundation (NSF)-MRSEC program (DMR-2011750) and the Gordon and Betty Moore Foundation’s EPIQS initiative (grant number GBMF9064). We thank Dr.~Yingkai Huang for providing ZrSiS crystals for low magnetic field measurements. 
\end{acknowledgments}

\nocite{*}

\bibliography{ZrSiS}

\clearpage

\newcommand{\beginsupplement}{%
        \setcounter{table}{0}
        \renewcommand{\thetable}{S\arabic{table}}%
        \setcounter{figure}{0}
        \renewcommand{\thefigure}{S\arabic{figure}}%
    }
\beginsupplement

\onecolumngrid
\section{Supplemental Material}
\subsection{Strain cell set-up}
Fig. S1 shows the completed sample mounting set-up for the ZrSiS sample measured in this work. The strain cell used was the Razorbill CS100. The bar-shaped sample (thickness $\sim120~\mu$m) is shown to be glued between the titanium sample plates using Araldite glue with a 24 hour curing time. In order to prevent an electrical short, four insulated 70 $\mu$m copper wires were placed between the sample and each sample plate right after the glue was applied. The thinner 25 $\mu$m gold wires were attached to the sample on the titanium-gold sputtered contact pads using DuPont 4929N silver conductive paint. Using the four-point probe method the sample resistance is measured. The wires are attached to the Razorbill WP100 wiring platform on the left, and a piece of double-sided tape on the right.

\begin{figure}[h]
\includegraphics[scale=0.4]{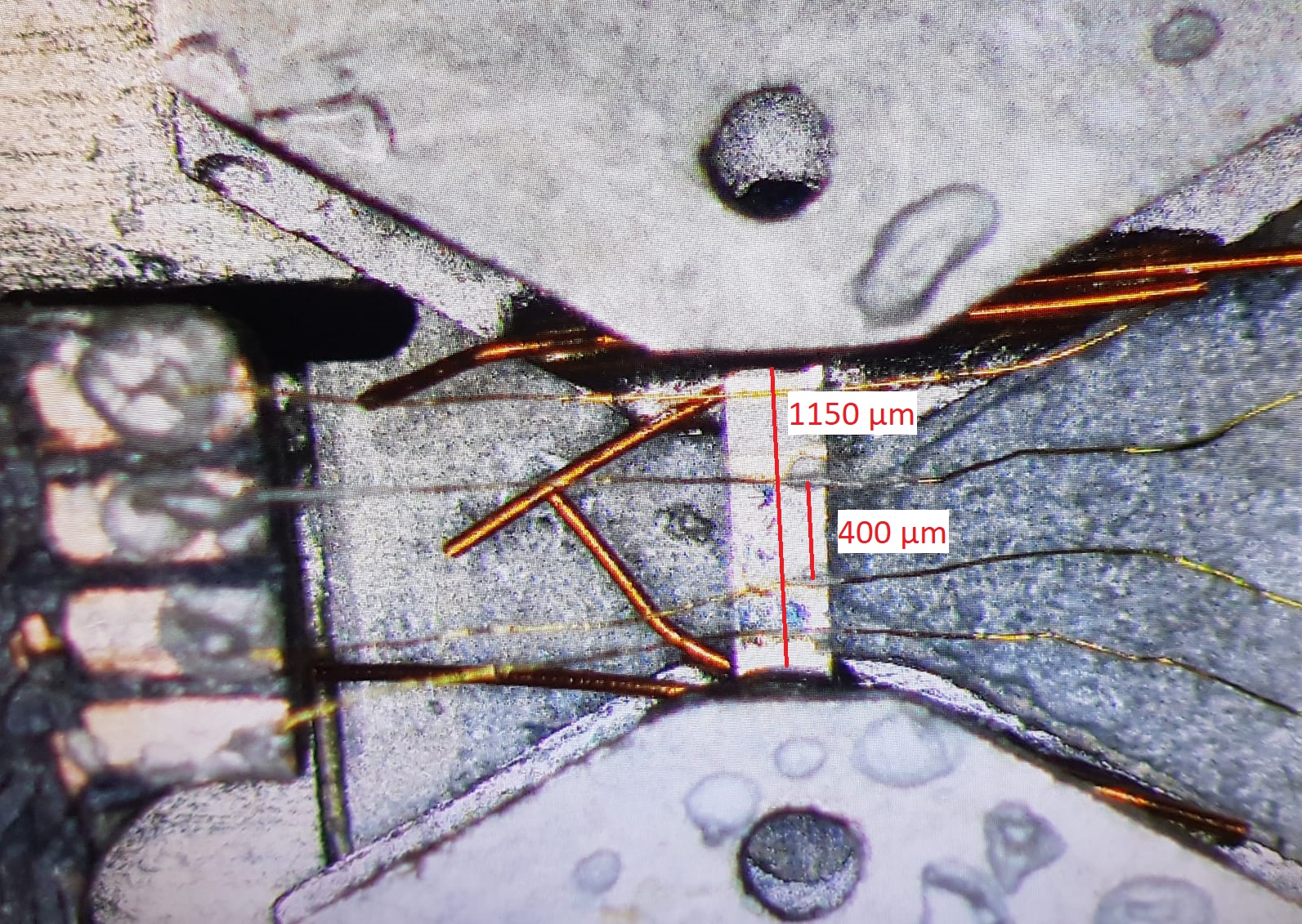}
\centering
\caption{The ZrSiS single crystal mounted on the Razorbill CS100 strain cell, with electrical wires arranged according to the four-point probe method. The crystal's $ab$ plane faces up and the magnetic field is applied along the $c$ axis. The longer red bar indicates the shortest distance $L$ between the two glued ends of the sample, while the shorter red bar indicates the distance between the two voltage contacts.}
\label{fig:setup}
\end{figure}

\newpage
\subsection{Magnetoresistance of strained ZrSiS relative to the zero strain measurement}
The difference in longitudinal resistance between the strained curves and the zero strain curve in Fig. 1 of the main text is defined as $\Delta R_{\textrm{xx}} = R_{\textrm{xx}, \varepsilon} - R_{\textrm{xx}, \varepsilon_0}$. This $\Delta R_{\textrm{xx}}$ is plotted against the magnetic field $B$ in Fig. S2, for the four different strain values corresponding to those of Fig. 1 of the main text. The oscillations in the $\Delta R_{\textrm{xx}}$ shown here indicate the tunability of the quantum oscillations through the use of uniaxial strain along the $a$~axis.

\begin{figure}[h]
\includegraphics[scale=0.6]{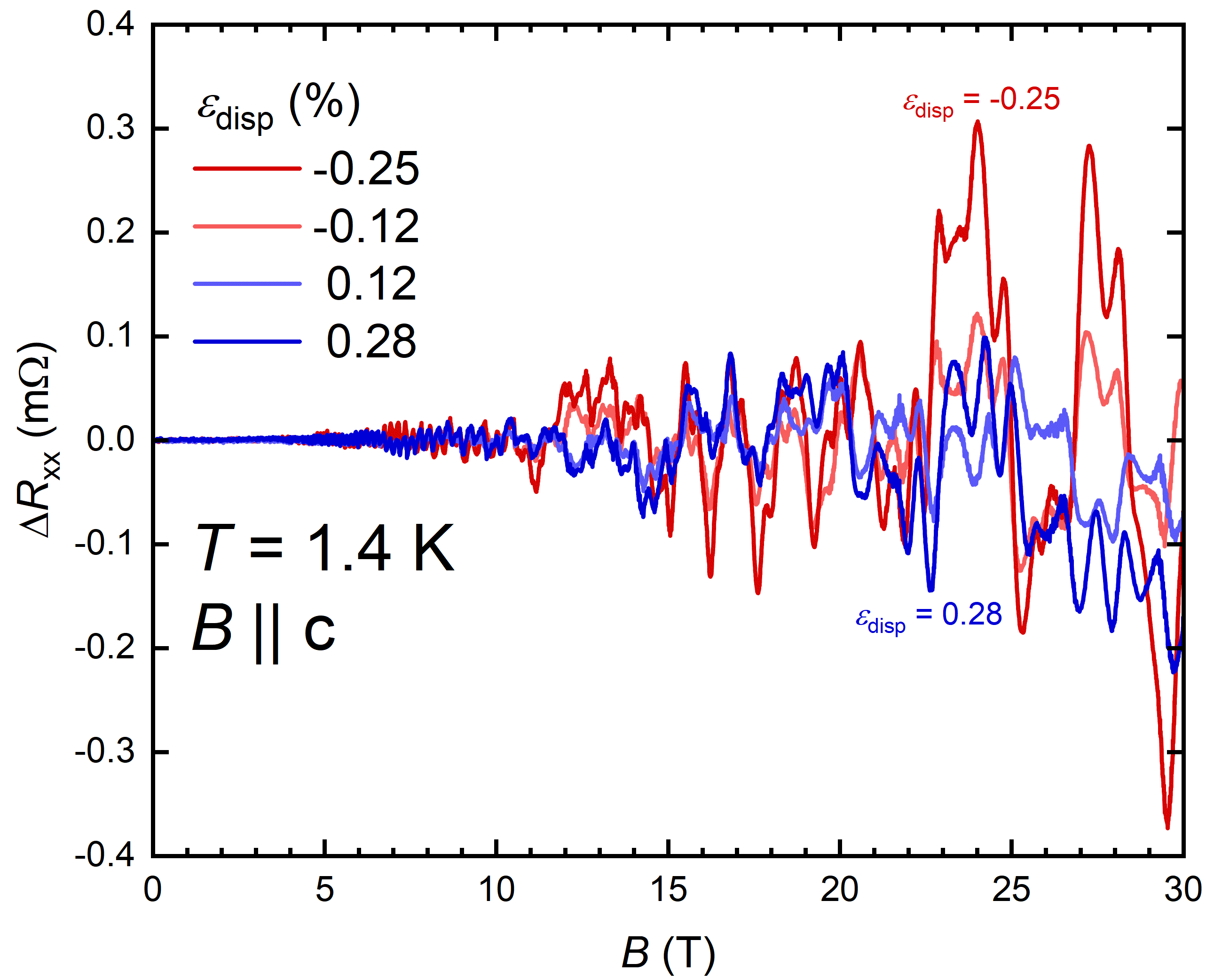}
\centering
\caption{Difference in longitudinal resistance $\Delta R_{\textrm{xx}}$ in m$\Omega$ compared to the zero strain curve shown in Fig. 1 of the main text, against the applied magnetic field $B$ in Tesla for negative and positive strain at 1.4~K.}
\label{fig:MRdiff}
\end{figure}

\newpage
\subsection{Magnetoresistance curves of temperature-dependent strain measurements}
The longitudinal magnetoresistance acquired by sweeping the magnetic field parallel to the $c$ axis from 0 to 30 T, at six different temperatures and three different strain values, is displayed in Fig. S3. Negative strain means compression, positive strain means tension. The three sets of strain curves have been offset by 1 m$\Omega$ for clarity. Curves with higher oscillation amplitudes correspond to lower-temperature data. Note the large increase in background magnetoresistance of the curves taken under tensile strain at 3.2, 4.2 and 6.0~K, possibly as a result of non-elastic deformation. These latter three curves were not used in the analysis of the FFT spectrum of ZrSiS under strain in the main text.

\begin{figure}[h]
\includegraphics[scale=0.6]{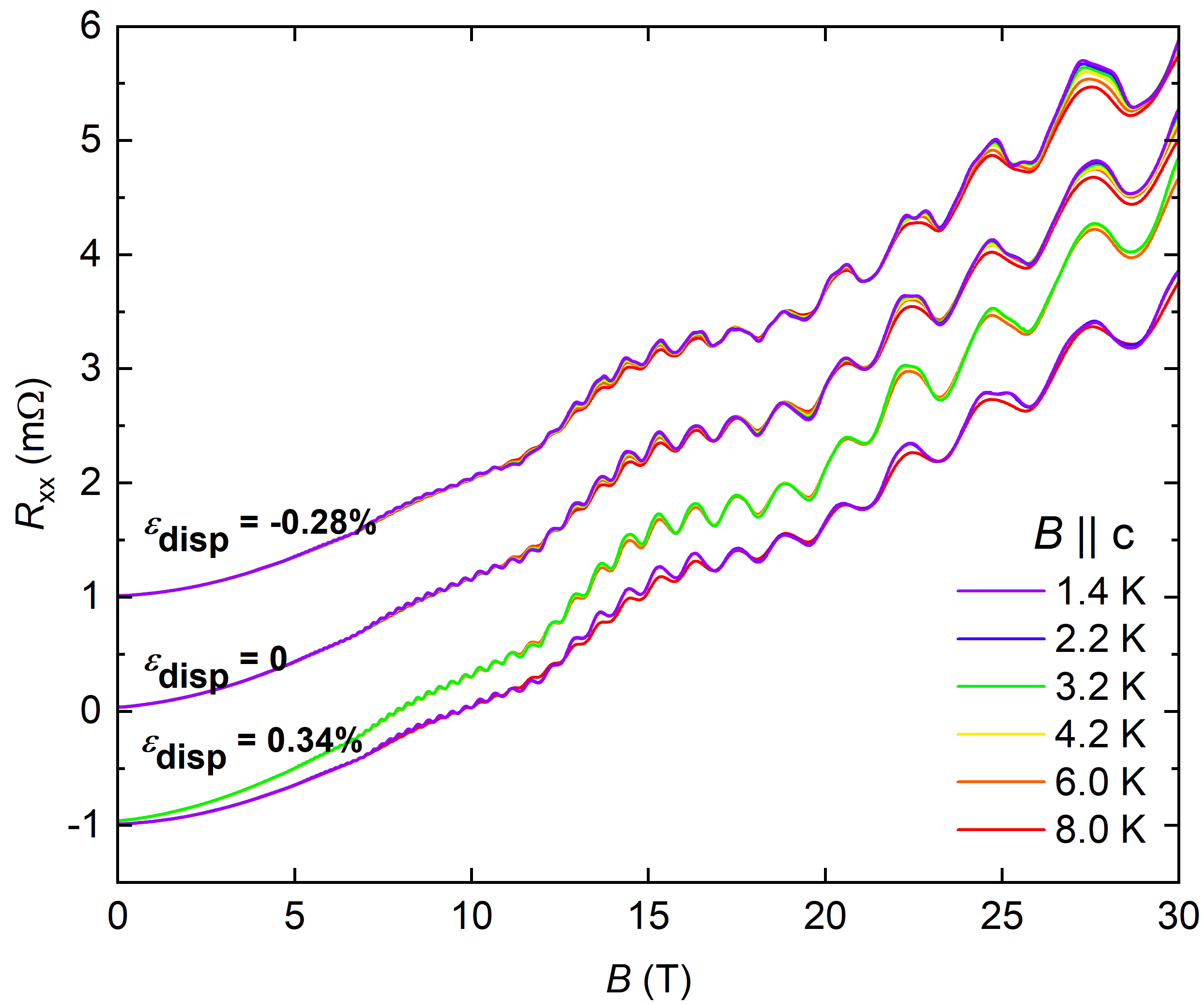}
\centering
\caption{Magnetoresistance at varying temperature (1.4 to 8.0~K) and different strain values. These curves are used to produce Fig. 2 of the main text. The curves for the different strains have been offset for clarity. Curves with higher oscillation amplitudes correspond to lower-temperature data.}
\label{fig:raw}
\end{figure}

\newpage
\subsection{FFT plot of the data shown in Fig.~1 of the main text}
The FFT plot of Fig. S4 shows that, just like for the $T$-dependent run shown in Fig.~2 of the main text, uniaxial strain along the $a$~axis has a systematic effect on the frequency peak structure. Again a weakening of the peak structure is observed upon changing from compressive to tensile strain.

\begin{figure}[h]
\includegraphics[scale=0.6]{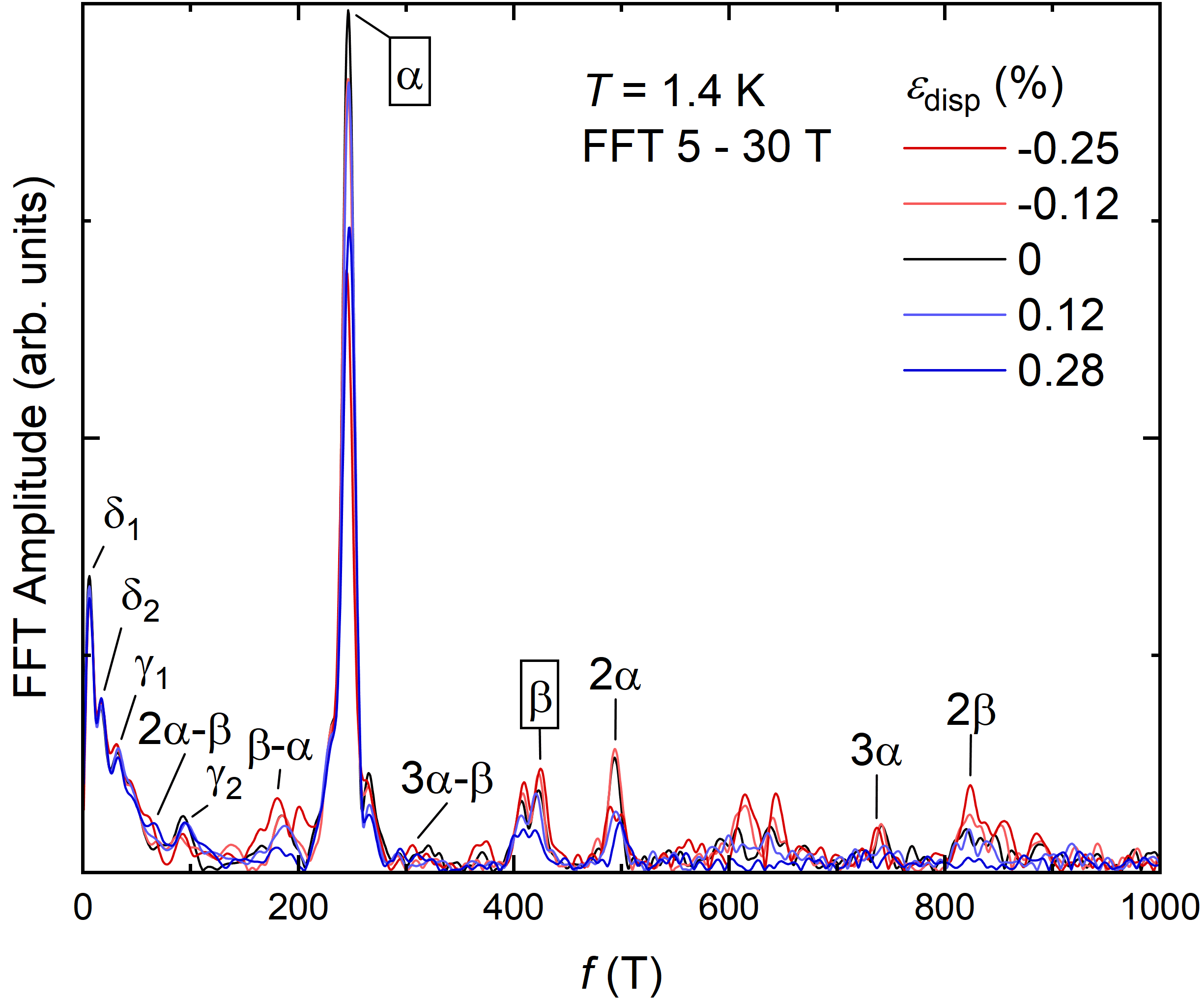}
\centering
\caption{FFT plot of the data shown in Fig.~1 of the main text. The FFTs were taken in the same way as described for Fig.~2 in the main text, with first derivatives and a magnetic field range of $5 - 30$~T using the Hann window function.}
\label{fig:FFTs}
\end{figure}

\newpage
\subsection{Extraction of cyclotron mass using the Lifshitz-Kosevich fit}
The cyclotron mass for a quasiparticle orbit frequency is obtained by fitting the temperature evolution of the FFT peak amplitude to the thermal damping factor of the Lifshitz-Kosevich formula $R_T = X/\sinh{(X)}$, where $X = 14.69 m_cT/\langle B \rangle$. The value for the average field here is defined as $\langle B \rangle = 2/(B_{min}^{-1} +~B_{max}^{-1})$.\\
Fig. S5 shows the fitting process of the compressed $\alpha$ orbit and zero strain $\beta$ orbit data for different field ranges and therefore different $\langle B \rangle$. Note that for the zero strain $\beta$ orbit data the small FFT amplitudes make taking the proper peak value difficult and lead to the relatively large $m_c$ fitting error shown in Fig. 3 of the main text.

\begin{figure}[h]
\includegraphics[scale=0.18]{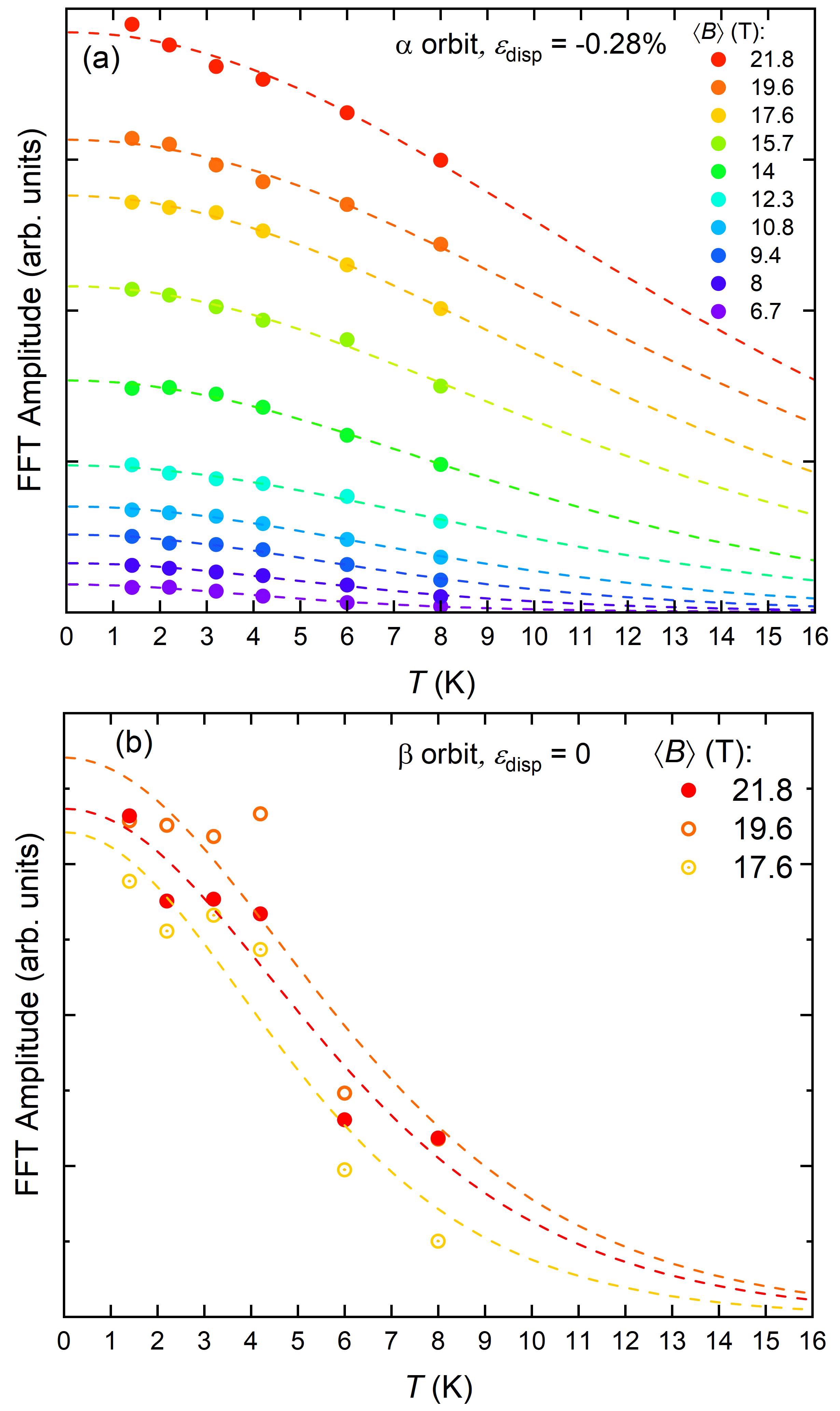}
\centering
\caption{Temperature dependence of the FFT amplitude of the (a) $\alpha$ pocket at $0.28~\%$ compression and (b) $\beta$ pocket at zero strain for different field ranges. The different ranges result in different average fields with $\langle B \rangle = 2 / (1/B_{max} + 1/B_{min})$ and are indicated by the different colors (in (a), curves from top to bottom correspond to descending $\langle B \rangle$ values). Ranges were taken with equal width in 1/$B$. Dashed lines are Lifshitz-Kosevich fits used to extract the cyclotron mass $m_{c,\alpha}$ of the orbit. These fits are used to produce Fig. 3 of the main text. Other cyclotron masses were obtained in the same way.}
\label{fig:LK}
\end{figure}

\newpage
\subsection{Comparing of observed quasiparticle orbit frequencies to literature}
Table S1 lists each observed quasiparticle orbit frequency in this work and compares their values with those listed in the work of M\"uller \emph{et al.} (Ref.~\cite{Mull20}) and the frequencies obtained from their density functional theory (DFT) calculations.

\begin{table}[h]
\centering
\caption{\label{tab:freqs}%
The identified orbits and their observed frequencies, compared to the frequencies observed by M\"uller \emph{et al.} (Ref.~\cite{Mull20}) and the frequencies obtained from their DFT calculations. The fundamental frequencies are highlighted in bold. Harmonics such as $2\alpha$ and $2\beta$ have been omitted.
}
\begin{tabular}{| l c c c |} 
 \hline
 Orbit & $f$(T) & $f_{\textrm{M\"uller}}$(T) & $f_{\textrm{DFT}}$(T) \\ [0.5ex] 
 \hline\hline
 $\delta_1$ & \textbf{7} & 8 & 7 \\ 
 \hline
 $\delta_2$ & \textbf{16} & 16 & 17 \\
 \hline
 $\gamma_1$ & \textbf{31} & 22 & 57 \\
 \hline
 $2\alpha-\beta$ & 72 & 60 & - \\
 \hline
 $\gamma_2$ & \textbf{88} & 90 & 156 \\
 \hline
 $\beta-\alpha$ & 173 & 180 & - \\ 
 \hline
 $\alpha$ & \textbf{245} & 240 & 240 \\
 \hline
 $3\alpha-\beta$ & 317 & 300 & - \\
 \hline
 $\beta$ & \textbf{418} & 420 & 518 \\
 \hline
 $4\alpha-\beta$ & - & 540 & - \\ 
 \hline
 $2\beta-\alpha$ & 591 & 600 & - \\ 
 \hline
 $\alpha+\beta$ & 663 & - & - \\ %[1ex]
 \hline
\end{tabular}
\end{table}

\newpage
\subsection{Elastic properties of ZrSiS}
Poisson's ratios for ZrSiS are determined in the same way as in Schindler \emph{et al.} (Ref.~\cite{Schi20}) using Voigt notation. The values for the stiffness tensor $C$ of ZrSiS calculated by de Jong \emph{et al.} (Ref.~\cite{Jong15}) are $C_{11} = 255$, $C_{12} = 69$, $C_{13} = 77$, $C_{33} = 187$, $C_{44} = 84$ and $C_{66} = 122$ GPa, at zero pressure and zero temperature. Tension along the $a$ axis results in compression along the $b$ axis, with Poisson's ratio
\begin{align}
    \nu_{21} = \frac{C_{12}C_{33}-C^2_{13}}{C_{11}C_{33}-C^2_{13}}
\end{align}
and also results in compression along the $c$ axis, with Poisson's ratio
\begin{align}
    \nu_{31} = \frac{(C_{11}-C_{12})C_{13}}{C_{11}C_{33}-C^2_{13}}.
\end{align}
This results in $\nu_{21} = 0.17$ and $\nu_{31} = 0.34$.\\
In its unstrained tetragonal state, the lattice parameters of ZrSiS are $a = b = 3.556$ \AA\ and $c = 8.180$ \AA. $a = b = 3.542$~\AA\ and $c = 8.049$ \AA. These are the values obtained by M\"uller \emph{et al.} (Ref.~\cite{Mull20}) by using X-ray diffraction on samples grown with the same method, by the same group (Schoop lab at Princeton University). In this study, the sample is strained up to $\varepsilon_{\textrm{disp}}$~$=$~$-0.28~\%$ compression and $\varepsilon_{\textrm{disp}}$~$=$~$0.34~\%$ tension along the $a$~axis. See Table S2 below for changes in lattice parameters and volume with strain. The percentual changes in $c/\sqrt{ab}$ and their corresponding $\mu_q$ values with tensile strain show a trend that closely matches that of $c/a$ and $\mu_q$ with hydrostatic pressure in Gu \emph{et al.} (Ref.~\cite{Gu19}). We use $c/\sqrt{ab}$ instead of $c/a$ to account for the fact that $a~\neq~b$ under uniaxial strain.

\begin{table}[h]
\centering
\caption{\label{tab:lattparams}%
The effect of uniaxial strain along the $a$~axis on the lattice parameters and unit cell volume of ZrSiS. The negative compressive and positive tensile strain values from the main text have been used, together with the calculated Poisson's ratios.
}
\begin{tabular}{| c c c c c c c c |} 
 \hline
 $\varepsilon_{\textrm{disp}}$(\%) along $a$ & $a$(\AA) & $b$(\AA) & $c$(\AA) & V(\AA$^3$) & $\Delta$V(\%) & $c/\sqrt{ab}$ & $\Delta (c/\sqrt{ab})(\%)$\\ [0.5ex] 
 \hline\hline
 0 & 3.542 & 3.542 & 8.049 & 100.981 & 0 & 2.272 & 0\\
 \hline
 -0.28 & 3.532 & 3.544 & 8.057 & 100.853 & -0.13 & 2.277 & 0.21\\
 \hline
 0.34 & 3.554 & 3.540 & 8.040 & 101.153 & 0.17 & 2.267 & -0.25\\
 \hline
\end{tabular}
\end{table}

\newpage
\subsection{Band structure calculations of ZrSiS under strain}
The Fermi surfaces of four structures with different sets of lattice parameters were obtained with density functional theory (DFT). The results are based on the all-electron full-potential augmented plane-wave (FLAPW) formalism, as implemented in the SPEX code (Friedrich \emph{et al.}, Ref.~\cite{Friedrich2010}). We used the generalized-gradient approximation (GGA) in the PBE parametrization (Perdew \emph{et al.}, Ref.~\cite{Perdew1996}). Spin-orbit coupling was included in the calculations in a second-variation scheme. 
The bases for the calculations were constructed by setting the maximal angular momentum $l_{max}$ as 11, 8, and 8 for the muffin-tin parts of Zr, Si, and S, respectively, and $k_{max}=3.99$\,bohr$^{-1}$ for the interstitial part.
The Brillouin zone integration was performed on a $14\times14\times14$ Monkhorst-Pack $\mathbf{k}$-point mesh. The data for the Fermi surface plots were generated by the Wannier90 code~\cite{wannier90}. We used Wannier90 to interpolate the bands around the Fermi level at a plane with $k_z=0.5$ rlu, and $k_{\{x,y\}} \in [-0.5,0.5]$ rlu, with $101 \times 101$ $\mathbf{k}$-point mesh, where ``rlu'' refers to ``relative (reciprocal) lattice unit''. The Wannier90 interpolated bands were subsequently interpolated by cubic interpolation $4$ times and $10$ times (thus $4 \times (101 \times 101)$ or $10 \times (101 \times 101)$ $\mathbf{k}$-point mesh after cubic interpolation) for both panels of Figure S7.

The lattice parameters of the four structures are shown in Table S3. The internal positions were kept fixed for the four structures as: $Z_{Zr} = 0.7308$ and $Z_S = 0.6225$. \\

\begin{table}[h]
\centering
\caption{\label{tab:DFT}%
The four structures with different sets of lattice parameters used to obtain the different Fermi surfaces of ZrSiS. Structure 1 uses values obtained from the Materials Project (Ref.~\cite{MatProj2013}) and structure 2 uses values obtained by M\"uller \emph{et al.} (Ref.~\cite{Mull20}). Structures 3 and 4 are the values from M\"uller under compression and tension, respectively, detailed in subsection G.}

\begin{tabular}{| c c c c c |}
\hline
 & Mat. Proj. & M\"uller & Compressed & Tensioned \\
 & (structure 1) & (structure 2) & (structure 3) & (structure 4)\\
\hline\hline
$a$ (\AA) & 3.554 & 3.542 & 3.532 & 3.554 \\
\hline
$b$ (\AA) & 3.554 & 3.542 & 3.544 & 3.540 \\
\hline
$c$ (\AA) & 8.118 & 8.049 & 8.057 & 8.040 \\
\hline
\end{tabular}
\end{table}

\newpage
Fig. S6 shows the calculated DFT-based band structures of ZrSiS for the four structures mentioned in Table S3. Fig. S7(a) shows the corresponding 2D Fermi surfaces in the Z-R-A~plane in reciprocal lattice units (rlu). Fig. S7(b) shows the MB gap for each structure more closely. Note that there is no significant change in MB gap size: the Fermi surfaces slightly shift relative to each other, but the gap size stays the same. The zero-strain results of the calculations are also in excellent agreement with those performed by M\"uller \emph{et al.} and Pezzini \emph{et al.} (Ref.~\cite{Mull20} and Ref.~\cite{Pezz18})

\begin{figure}[h]
\includegraphics[scale=0.11]{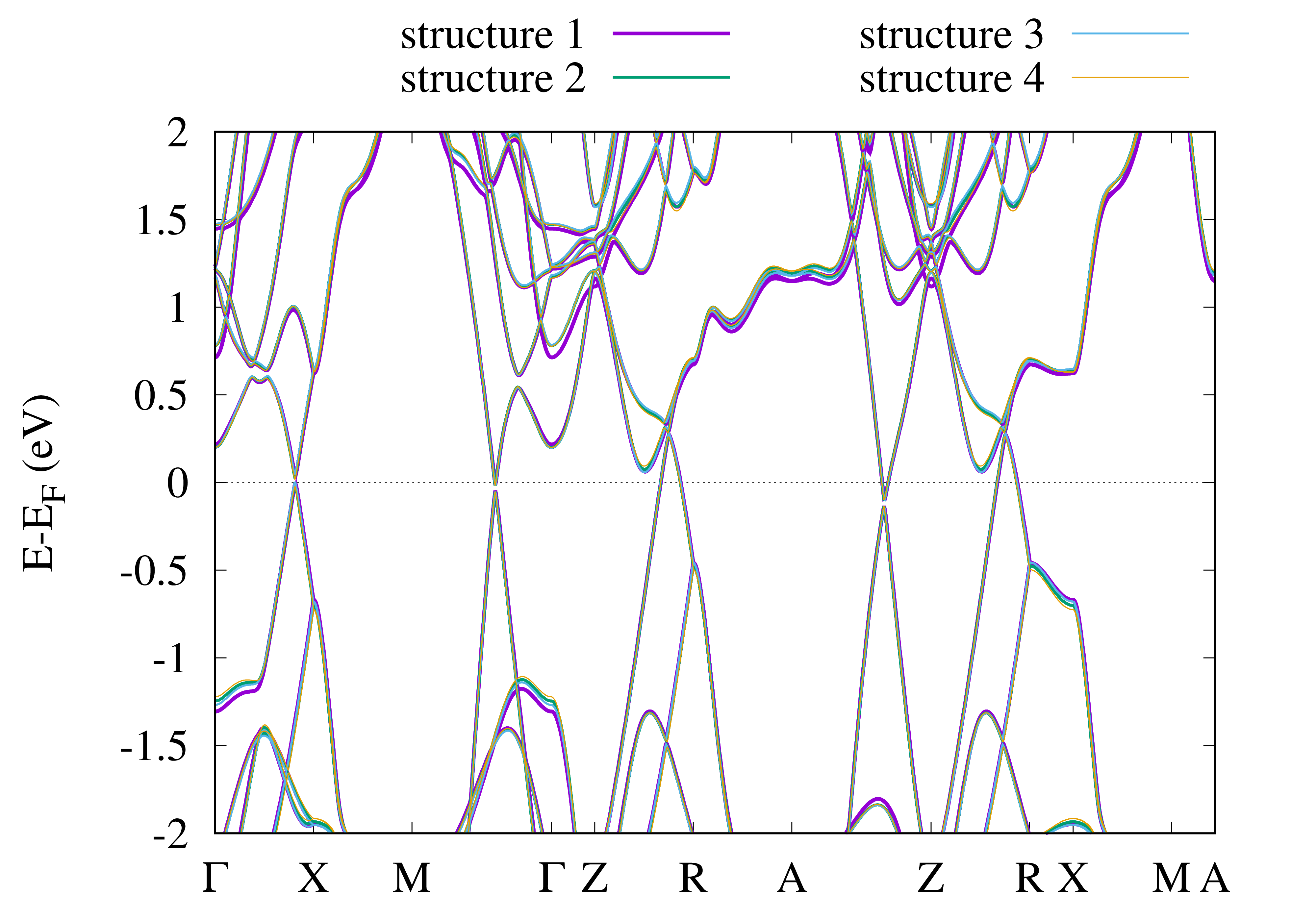}
\centering
\caption{Calculated DFT-based band structures of ZrSiS for the four structures mentioned in Table S3.}
\label{fig:DFTbands}
\end{figure}

\begin{figure}[h]
\includegraphics[scale=0.51]{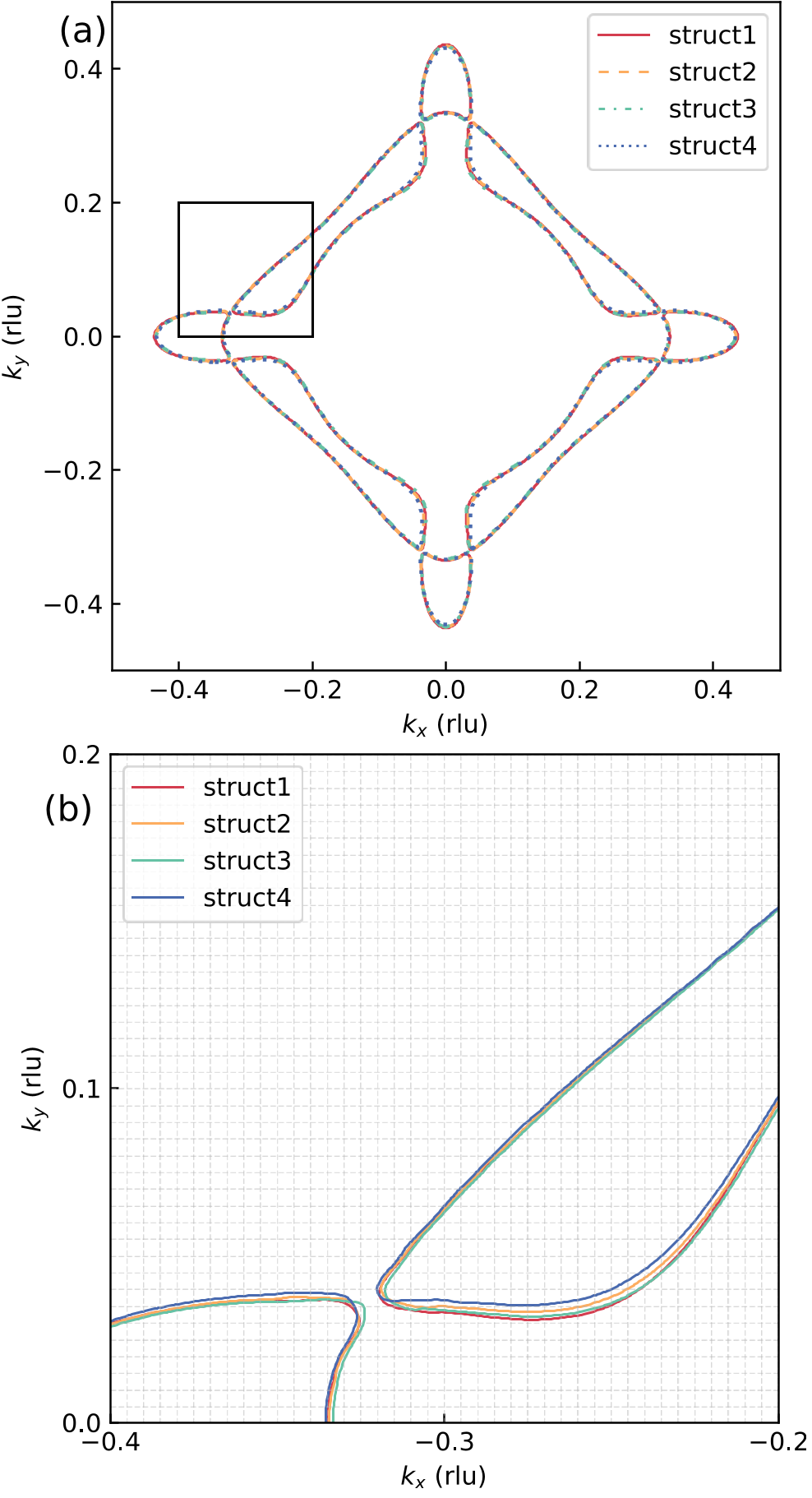}
\centering
\caption{The 2D Fermi surfaces in the Z-R-A~plane corresponding to the band structures of Fig. S6 are shown in (a). The area enclosed by the black square is displayed larger in (b).}
\label{fig:DFT2DFS}
\end{figure}

\end{document}